\documentclass[twocolumn]{aastex631mod}
\usepackage{booktabs} 
\usepackage{dcolumn}  
\usepackage[UKenglish]{isodate}

\providecommand{\myeol}{\\}
\providecommand{\dubious}{$\dagger$}
\providecommand{\knownbinary}{b}
\providecommand{\CI}{90\% CI}

\providecommand{\gaia}{Gaia}
\providecommand{\gdr}[1]{Gaia DR{#1}}

\providecommand{\parallax}{\ensuremath{\varpi}}
\providecommand{\propm}{\ensuremath{\mu}}
\providecommand{\rv}{\ensuremath{v_r}}
\providecommand{\vt}{\ensuremath{v_t}}
\providecommand{\gmag}{\ensuremath{G}}
\providecommand{\bprp}{\ensuremath{\rm BP-RP}}
\providecommand{\mg}{\ensuremath{M_G}}
\providecommand{\teff}{\ensuremath{T_{\rm eff}}}

\providecommand{\dphlimit}{\ensuremath{d_{\rm lim}}}

\providecommand{\tph}{\ensuremath{t_{\rm ph}}}
\providecommand{\dph}{\ensuremath{d_{\rm ph}}}
\providecommand{\vph}{\ensuremath{v_{\rm ph}}}
\providecommand{\tphmed}{\ensuremath{t_{\rm ph}^{\rm med}}}
\providecommand{\dphmed}{\ensuremath{d_{\rm ph}^{\rm med}}}
\providecommand{\vphmed}{\ensuremath{v_{\rm ph}^{\rm med}}}

\providecommand{\deg}{\ensuremath{^\circ}}

\providecommand{\mas}{\ensuremath{\rm mas}}
\providecommand{\kms}{\ensuremath{\rm km\,s^{-1}}}
\providecommand{\maspyr}{\ensuremath{\rm \mas\,yr^{-1}}}
\providecommand{\Msol}{\ensuremath{\rm M_\odot}}

\defcitealias{2015A&A...575A..35B}{paper~1} 
\defcitealias{2018A&A...609A...8B}{paper~2} 
\defcitealias{2018A&A...616A..37B}{paper~3} 

\received{18 June 2022}
\revised{4 July 2022}
\accepted{13 July 2022 to ApJ Letters}
\shorttitle{Stellar encounters in \gdr{3}}
\shortauthors{Bailer-Jones}

\begin{document}

\title{Stars that approach within one parsec of the Sun:\\New and more accurate encounters identified in Gaia Data Release 3}
\author{C.A.L.\ Bailer-Jones}
\affiliation{Max Planck Institute for Astronomy, K\"onigstuhl 17, 69117 Heidelberg, Germany}

\begin{abstract}

Close encounters of stars to the Sun could affect life on Earth through gravitational perturbations of comets in the Oort cloud or exposure to ionizing radiation.
By integrating orbits through the Galactic potential, I identify which of 33 million stars in \gdr{3} with complete phase space information come close to the Sun. 
61 stars formally approach within 1\,pc, although there is high confidence in only 42 (two thirds) of these, the rest being spurious measurements or (in) binary systems. Most of the stars will encounter within the past or future 6\,Myr; earlier/later encounters are less common due to the magnitude limit of the \gaia\ radial velocities (RVs). Several close encountering stars are identified for the first time, and the encounter times, distances, and velocities of previously known close encounters are determined  more precisely on account of the significantly improved precision of \gdr{3} over earlier releases.
The K7 dwarf Gl 710 remains the closest known encounter, with an estimated (median) encounter distance of 0.0636\,pc (90\% confidence interval 0.0595--0.0678\,pc) to take place in 1.3\,Myr.
The new second closest encounter took place 2.8\,Myr ago: this was the G3 dwarf HD 7977, now 76\,pc away, which approached within less than 0.05\,pc of the Sun with a probability of one third. The apparent close encounter of the white dwarf
\object{UPM J0812-3529} is probably spurious due to an incorrect RV in \gdr{3}.

\end{abstract}


\section{Introduction} \label{sec:introduction}

Close encounters between stars can influence the evolution of stellar systems. Early in stars' lives, interactions can disrupt circumstellar disks, potentially influencing planet formation. 
Later in stars' lives, loosely bound reservoirs of comets, such as the Oort cloud around the Sun, can be perturbed by a close encounter. This can 
fling comets inward -- where they could impact planets -- or eject them out of the system, possibly leading to capture by the other star. The close passage of an active star, in particular one that goes supernova, could even jeopardize -- or perhaps assist -- the evolution of life on a planet.

Of particular interest is the stellar encounter history of the Sun. To identify past or future encounters we need to know stars' current positions and velocities. Extensive searches for encounters were therefore only possible starting with the publication of a hundred thousand parallaxes by Hipparcos in 1997, which increased by orders of magnitude two decades later with \gaia. These data led to the identification of many close encounters (e.g.\ \citealt{2001A&A...379..634G, 2017AstL...43..559B, 2018A&A...616A..37B, 2020A&A...640A.129W}) which fed into modelling how these encounters affected our Oort cloud (e.g.\ \citealt{2015MNRAS.454.3267F, 2019A&A...629A.139T, 2020MNRAS.491.2119W, 2022A&A...660A.100D}).

The present paper continues a study to discover and characterize close encounters, one that started with Hipparcos (\citealt{2015A&A...575A..35B}; paper 1), then \gdr{1} (\citealt{2018A&A...609A...8B}; paper 2) -- both complemented by non-\gaia\ data -- and most recently \gdr{2} (\citealt{2018A&A...616A..37B}; paper 3). 
Since the first \gaia\ data release, astrometry has been in abundance and the
comparative lack of relative radial velocities (RVs) has been the limiting factor of these studies, a complete reversal of the pre-Hipparcos situation.
\gdr{3} \citep{gdr3_release} now provides radial velocities for 34 million bright stars (99\% with $\gmag < 15.7$\,mag) with median uncertainties of 3.3\,\kms\ (central 90\% range 0.4--7.8\,\kms). This is nearly a five-fold increase in the number of sources with radial velocities in \gdr{2}, and constitutes the largest radial velocity survey to date.

Here I use these data to identify stars that approach within 1\,pc of the Sun.
Previous works by various authors have used a larger limit. 
But as the average spacing\footnote{By counting the number of stars within some distance we can compute the spacing they would have if they were on a regular cubic lattice. From the \gaia\ Catalogue of Nearby Stars \citep{2021A&A...649A...6G} we get values of 2.23 to 2.33\,pc depending on the distance used and whether correcting for incompleteness. Using all sources in \gdr{3} we get 1.95--2.37\,pc for various distances out to 100\,pc.} between stellar systems in the solar neighbourhood (a few tens of pc) is about 2.2\,pc, 
and the closest stellar neighbour to the Sun -- the Alpha Centauri system -- is currently just 1.3\,pc away, 1\,pc seems a more meaningful upper limit. The Oort cloud extends to perhaps 0.25--0.50\,pc from the Sun \citep[e.g.][]{2012Icar..217....1B}, and even massive stars passing much further from the Sun than this are expected to have only a small effect on the Oort cloud \citep[e.g.][]{2022MNRAS.tmp.1749B}.

\section{Data} \label{sec:data}

\gdr{3} contains 33\,653\,049 sources with complete six-dimensional phase space coordinates (three positions and three velocities).
Of these, 29\,947\,046 have a parallax signal-to-noise ratio (SNR), {\tt parallax\_over\_error}, greater than 5 (prior to adjusting the parallax zeropoint). I adopt this as the main sample to search for encounters. Less precise parallaxes correspond to less precise encounter parameters, such that even if the median perihelion distance (\dphmed) were small, its probability distribution would be broad.
I nonetheless applied the procedure described in section~\ref{sec:method} to the 3\,311\,812 sources with 
$0 <$\,{\tt parallax\_over\_error}\,$< 5$, but none reached $\dphmed<1$\,pc.
I did not consider at all the 394\,191 sources with {\tt parallax\_over\_error}\,$\leq0$. Although it is possible to infer distances for sources with negative parallaxes (e.g.\ \citealt{2015PASP..127..994B}), their distance posteriors are prior-dominated and generally very broad, so the encounter parameters remain very imprecise.

Once selected, I corrected the parallax zeropoint of all the sources in the main sample using the procedure described by \cite{2021A&A...649A...4L}. This corresponds to increasing the raw parallaxes by 0.006 to 0.058\,mas (central 98\% range), the exact value depending on magnitude (as well as colour and sky position).
Other analyses have suggested that the parallaxes may have to be increased further
for the brighter stars in \gaia, by up to 0.015\,mas (e.g.\ \citealt{2021AJ....161..214Z, 2021ApJ...911L..20R}).
This would have little impact on my encounter results, however:
Although the raw parallaxes of the main sample cover a broad range, 0.12 to 7.9\,mas (central 98\% range), 
a star will generally have to be closer than about 1\,kpc
in order to have a high probability of approaching within 1\,pc of the Sun, 
such that any residual error in the parallax zeropoint is small compared to the parallax. 

There is some evidence that some of the parallax uncertainties in \gdr{3} are underestimated (e.g.\ \citealt{2022A&A...657A.130M}), but I have not inflated them in this study.

An observed radial velocity is often not equal to the true radial velocity for a number of physical reasons beyond measurement or calibration errors. The gravitational redshift increases the observed radial velocity by about 0.6\,\kms\ for main sequence stars (less for giants), although this is offset by convective blueshifts that  reduce the radial velocity by a similar size: As discussed by \cite{2002A&A...390..383G}, the net effect is that the observed minus true radial velocity ranges from about $-0.4$\,\kms\ for F stars to $+0.4$\,\kms\ for K dwarfs. These offsets are generally less than the \gdr{3} radial velocity uncertainties, and as the specific correction depends on properties that are not well determined for most stars, no correction has been made.

\section{Method}\label{sec:method}

Once we have the phase space coordinates and specify the Galactic gravitational potential, the orbits of stars relative to the Sun can be computed by numerical integration. From this we can determine the time, distance, and velocity of closest approach. An accurate numerical integration for all 30 million stars is not only 
time-consuming, it is also unnecessary, because the vast majority of the stars will never come anywhere near the Sun. 
Moreover, the uncertainties in (and covariances between) the phase space coordinates must be taken into account.
I therefore adopt the following procedure for each star, which is an improvement over the procedure used in papers 1--3 (but the method of integration is unchanged).

\begin{enumerate}

\item Resample the phase space coordinates 50 times assuming the mean and covariances from \gdr{3} represent a six-dimensional Gaussian probability distribution over the coordinates, to give 50 ``surrogate" stars.

\item Find the closest approach between each surrogate and the Sun under the assumption of nonaccelerated motion, i.e.\ neglecting gravity. This linear motion approximation (LMA) has a simple analytic solution (section 3.2 of 
\citetalias{2015A&A...575A..35B}). 

\item If any of the 50 surrogates from the previous step approaches within \dphlimit\,=\,7.07\,pc of the Sun, numerically integrate the orbits of all 50 surrogates through the Galactic potential. This uses 50 time steps distributed uniformly over the interval $[0, 2\tph^{\rm LMA}]$, where $\tph^{\rm LMA}$ is the time of closest approach computed from the LMA (for that surrogate).

\item If any of the 50 surrogates from the previous step approaches within \dphlimit\,=\,7.07\,pc of the Sun, do a higher resolution numerical integration using 1000 surrogates each integrated for 500 time steps.

\item Determine whether the median (over the 1000 surrogates; \dphmed) encounter distance is below 1\,pc. 

\end{enumerate}
Only the stars that make it through the final step are reported here.
Using the distributions over the perihelion time (\tph, signed), distance (\dph, non-negative), and velocity (\vph, non-negative), I compute their medians as well as their 5\% and 95\% quantiles, the latter two providing the 90\% central confidence interval (CI) for each parameter.

\dphlimit\ was chosen to be large enough to make it likely that the selections in steps 3 and 4 include all encounters that will come within 1\,pc when integrated at high resolution in step 5. If \dphlimit\ is lowered to 5\,pc 
then in fact we miss two encounters.\footnote{$(7.07/5)^2 \simeq 2$. The expected number of encounters within a distance $d$ scales as $d^2$ \citep{2018A&A...609A...8B}.}
It is therefore possible that a few more encounters would be found if we used a larger \dphlimit, although this gets increasingly unlikely.

The LMA in step 2 is reasonably accurate in many cases, and good enough for the first liberal selection, because many stars travel only a short distance from their current location to perihelion, so their paths are relatively unaccelerated by the Galactic potential. 
The gravitational potential used in steps 3 and 4 is a simple, three-component axisymmetric model described in section 3.3 of 
\citetalias{2015A&A...575A..35B}. Because the accelerations are not large, the results are not very sensitive to the exact potential or location of the Sun in the Galaxy. As discussed in section 5.2 of 
\citetalias{2015A&A...575A..35B}, close encounters of a star on its journey to perihelion with other individual stars will hardly affect its orbit in the vast majority of cases. We may, therefore, reasonably adopt a smooth potential.


Light travel times have been neglected in the computations. This makes the 
inferred encounter times too late (too positive) by an amount roughly equal to the light travel time, 
which in the worst case is about 1\,kyr. 
The fractional error this introduces in the encounter time is approximately the ratio of the radial velocity to the speed of light, which is of order $4\times 10^{-4}$.
As we will see, this is much smaller than the uncertainties arising from the \gaia\ measurements.

For a very close encountering star, or one with large measurement uncertainties, it is possible that the orbits of the surrogates -- which represent these uncertainties -- pass the Sun on opposite sides. 
If we then calculated the average of the {\em signed Cartesian coordinates} of these surrogates at perihelion, this average could be arbitrarily close to zero, on account of the cancelling of positive and negative coordinates (surrogates passing on either side).
But this is clearly not a useful measure of the average perihelion {\em distance}.
We must instead look at the distribution of the (non-negative) perihelia distances over the surrogates, and then average those. 
This is what I do in this work (and did in papers 1--3), using the quantiles of the distribution to characterize the distribution. 
Not all studies do this \citep[e.g.][]{2022arXiv220611047D}, 
and the average signed coordinate will always produce a lower ``distance" than the average of the non-negative distances.
In principle a similar problem can also occur with the encounter time or velocity if the radial velocity is near zero. The encounter times I compute are signed, but we can see from the distribution (see \citetalias{2015A&A...575A..35B})  -- summarized by the \CI\ -- whether some times have a different sign. This does not occur for any of the encounters in this study.

\section{Results}\label{sec:results}

\subsection{Overall results}\label{sec:results:overall}

Of the 29\,947\,046 stars subject to the LMA analysis in step 2 of the procedure in section~\ref{sec:method}, 11\,199 had at least one surrogate with $\dph<7.07$\,pc and so were processed in step 3. Of those, 6407 had at least one surrogate with $\dph<7.07$\,pc in the low resolution orbit integration and so were processed in step 4. Of those, 1892 had a median encounter distance (\dphmed) below 5\,pc following the high resolution orbital integration. 61 have $\dphmed<1$\,pc; this is the sample that I examine in the rest of this paper.

\begin{figure*}[th]
\begin{center}
\includegraphics[width=0.75\textwidth, angle=0]{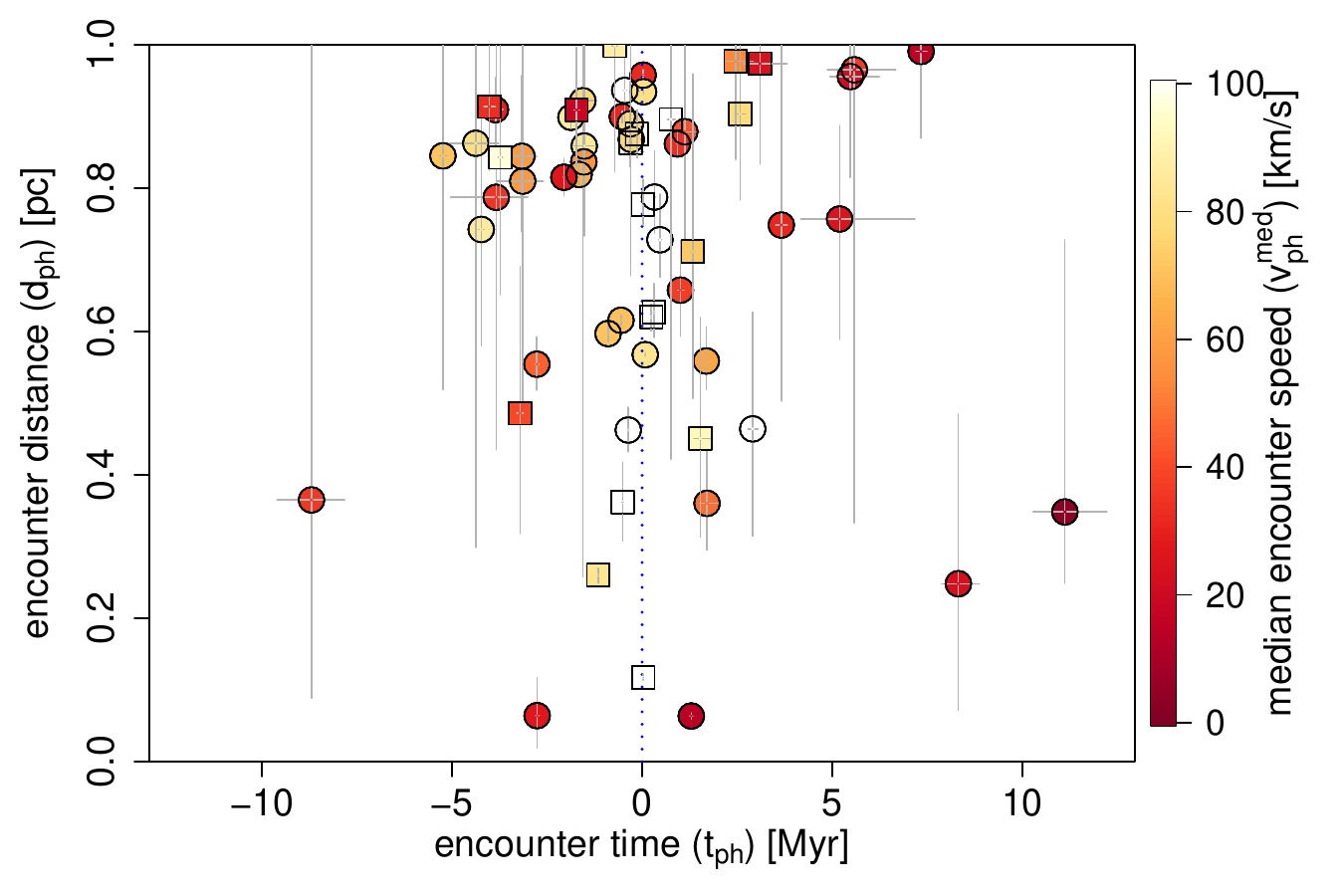}
\caption{Perihelion (encounter) times and distances for the 61 stars that have a median perihelion distance (\dphmed) below 1\,pc.  Negative times indicate past encounters, positive times future ones.
The circles/squares show the median of the perihelion time and distance distributions computed from the 1000 data resamples (surrogates); the error bars show the 5th and 95th percentiles. Circles denote good encounters; squares denote the questionable encounters, which are marked with a \dubious\ or \knownbinary\ in Table~\ref{tab:periStats}.
The colour of each point indicates the median encounter velocity; those faster than 100\,\kms\ are white.
\label{fig:dph_vs_tph}
}
\end{center}
\end{figure*} 

Figure~\ref{fig:dph_vs_tph} shows the encounter times, distances, and velocities of the 61 closest encounters. Most of the encounters take place within 6\,Myr of the present date. Larger absolute times correspond to stars that are currently further away and/or which have low velocities relative to the Sun. This ``time horizon" of the encountering stars arises from them having a characteristic maximum distance due to the magnitude limit of the main sample. This, combined with a characteristic velocity relative to the Sun, means most stars cannot travel for more than a few million years to/from their encounter and be visible now.

\begin{table*}[th]
\centering
\tiny{
\caption{Perihelion (encounter) parameters for all stars with a median perihelion distance (\dphmed) below 1\,pc, sorted by increasing \dphmed.
Columns 2, 5, and 8 are
$\tphmed$, $\dphmed$, and $\vphmed$ respectively. The columns labelled 5\% and 95\% are the bounds of the corresponding confidence intervals.
Columns 11--16 list the current parallax ($\parallax$, including the zeropoint offset), total proper motion ($\propm$), and radial velocity ($\rv$), along with their 1-sigma uncertainties.
Column 17 is the ratio of the current absolute radial velocity to the current transverse velocity, the latter computed as $4.74047\,\propm/\parallax$.
On the far left, \dubious\ indicates that the encounter is dubious for a variety of possible reasons, and \knownbinary\ indicates it is probably in a wide binary system according to \cite{2021MNRAS.506.2269E}, so may also not be reliable.
A machine readable version of this table reporting values to more
significant figures is available from \href{https://www.mpia.de/homes/calj/stellar_encounters_gdr3.html}{this URL}.
\label{tab:periStats}
}
\tabcolsep=0.14cm
\begin{tabular}{*{20}{r}}
\toprule
& & 1 & 2 & 3 & 4 & 5 & 6 & 7 & 8 & 9 & 10 & 11 & 12 & 13 & 14 & 15 & 16 & 17 \\
\midrule
& & \gdr{3} {\tt source\_id} & \multicolumn{3}{c}{$\tph$ [kyr]} &  \multicolumn{3}{c}{$\dph$ [pc]} &  \multicolumn{3}{c}{$\vph$ [\kms]} &
 $\parallax$ & $\sigma(\parallax)$ & $\propm$ & $\sigma(\propm)$ & $\rv$ & $\sigma(\rv)$ & $|\rv/\vt|$ \\
& &  & med & 5\% & 95\% & med & 5\% & 95\% & med & 5\% & 95\% & \multicolumn{2}{c}{\mas} & \multicolumn{2}{c}{\maspyr} &  \multicolumn{2}{c}{\kms} & \\
\midrule
\hphantom{\knownbinary} &  1 &  4270814637616488064 &     1292 &     1257 &     1334 &    0.064 &    0.060 &    0.068 &    14.4 &    14.0 &    14.8 &    52.43 &     0.02 &     0.43 &     0.02 &   -14.4 &     0.3 &   373 \myeol 
\hphantom{\knownbinary} &  2 &   510911618569239040 &    -2761 &    -2798 &    -2720 &    0.064 &    0.019 &    0.117 &    26.8 &    26.4 &    27.2 &    13.24 &     0.03 &     0.14 &     0.02 &    26.8 &     0.2 &   517 \myeol 
\dubious\hphantom{\knownbinary} &  3 &  5544743925212648320 &       29 &       28 &       30 &    0.117 &    0.113 &    0.122 &   374.3 &   359.7 &   387.1 &    89.54 &     0.02 &    71.70 &     0.02 &  -373.7 &     8.2 &    98 \myeol 
\hphantom{\knownbinary} &  4 &   213090546082530816 &     8316 &     7874 &     8874 &    0.248 &    0.072 &    0.485 &    23.0 &    21.5 &    24.3 &     5.07 &     0.01 &     0.45 &     0.01 &   -23.5 &     0.8 &    56 \myeol 
\knownbinary &  5 &  5571232118090082816 &    -1157 &    -1166 &    -1146 &    0.260 &    0.249 &    0.270 &    82.6 &    81.9 &    83.3 &    10.24 &     0.01 &     0.47 &     0.01 &    82.5 &     0.4 &   379 \myeol 
\hphantom{\knownbinary} &  6 &  5469802896279029504 &    11119 &    10281 &    12222 &    0.348 &    0.249 &    0.728 &     3.2 &     2.9 &     3.4 &    29.40 &     0.02 &     2.39 &     0.02 &    -2.7 &     0.2 &     7 \myeol 
\hphantom{\knownbinary} &  7 &  3372104035275483392 &     1706 &     1518 &     1982 &    0.360 &    0.295 &    0.441 &    47.6 &    41.0 &    53.5 &    12.03 &     0.04 &     0.71 &     0.04 &   -47.4 &     3.7 &   170 \myeol 
\dubious\hphantom{\knownbinary} &  8 &  3207963476278403200 &     -514 &     -529 &     -498 &    0.362 &    0.308 &    0.418 &   515.1 &   506.1 &   525.5 &     3.69 &     0.05 &     0.81 &     0.04 &   515.5 &     5.8 &   496 \myeol 
\hphantom{\knownbinary} &  9 &  3106500096597409792 &    -8695 &    -9599 &    -7823 &    0.365 &    0.088 &    1.027 &    36.2 &    32.7 &    40.1 &     3.09 &     0.01 &     0.48 &     0.01 &    36.9 &     2.2 &    50 \myeol 
\knownbinary & 10 &  4763293626627587840 &     1535 &     1361 &     1788 &    0.451 &    0.313 &    0.620 &    92.0 &    78.7 &   103.6 &     6.93 &     0.05 &     0.39 &     0.06 &   -91.1 &     7.4 &   341 \myeol 
\hphantom{\knownbinary} & 11 &   929788371508812288 &     -364 &     -384 &     -343 &    0.463 &    0.432 &    0.494 &   195.3 &   184.9 &   207.0 &    13.76 &     0.06 &     3.82 &     0.05 &   195.7 &     6.6 &   149 \myeol 
\hphantom{\knownbinary} & 12 &  6913732624445112832 &     2908 &     2782 &     3051 &    0.464 &    0.314 &    0.627 &   105.2 &   100.1 &   109.6 &     3.21 &     0.02 &     0.73 &     0.02 &  -104.1 &     2.9 &    96 \myeol 
\dubious\hphantom{\knownbinary} & 13 &  3118526069444386944 &    -3212 &    -3293 &    -3125 &    0.486 &    0.318 &    0.691 &    40.0 &    39.1 &    40.9 &     7.61 &     0.05 &     0.26 &     0.05 &    40.1 &     0.5 &   251 \myeol 
\hphantom{\knownbinary} & 14 &  6608946489396474752 &    -2768 &    -2819 &    -2712 &    0.554 &    0.518 &    0.593 &    44.8 &    44.0 &    45.7 &     7.95 &     0.01 &     0.72 &     0.01 &    43.8 &     0.5 &   102 \myeol 
\hphantom{\knownbinary} & 15 &  3295253979286613376 &     1687 &     1610 &     1782 &    0.559 &    0.519 &    0.607 &    61.8 &    58.4 &    64.7 &     9.38 &     0.02 &     0.67 &     0.02 &   -61.6 &     1.9 &   181 \myeol 
\hphantom{\knownbinary} & 16 &  1952802469918554368 &       83 &       82 &       83 &    0.568 &    0.566 &    0.570 &    83.2 &    82.9 &    83.5 &   141.94 &     0.02 &   201.01 &     0.02 &   -82.9 &     0.2 &    12 \myeol 
\hphantom{\knownbinary} & 17 &  5261593808165974784 &     -896 &     -907 &     -883 &    0.597 &    0.586 &    0.608 &    71.1 &    70.2 &    72.1 &    15.35 &     0.01 &     2.21 &     0.01 &    71.1 &     0.6 &   104 \myeol 
\hphantom{\knownbinary} & 18 &  3054509410098672000 &     -557 &     -561 &     -552 &    0.616 &    0.609 &    0.622 &    70.3 &    69.8 &    71.0 &    24.97 &     0.02 &     5.53 &     0.02 &    70.4 &     0.4 &    67 \myeol 
\dubious\hphantom{\knownbinary} & 19 &  3001468183198140800 &      242 &      238 &      246 &    0.620 &    0.601 &    0.638 &   885.7 &   873.0 &   896.8 &     4.56 &     0.03 &     1.66 &     0.03 &  -885.2 &     7.1 &   513 \myeol 
\dubious\hphantom{\knownbinary} & 20 &  4116451378388951424 &      311 &      294 &      329 &    0.627 &    0.592 &    0.667 &   264.8 &   254.9 &   273.4 &    11.91 &     0.33 &     4.22 &     0.17 &  -264.4 &     5.5 &   158 \myeol 
\hphantom{\knownbinary} & 21 &  3600338081985998080 &     1008 &      930 &     1109 &    0.658 &    0.594 &    0.727 &    37.4 &    34.0 &    40.4 &    25.97 &     0.08 &     3.19 &     0.11 &   -37.2 &     1.9 &    64 \myeol 
\dubious\hphantom{\knownbinary} & 22 &  4536673181955253504 &     1332 &     1278 &     1396 &    0.712 &    0.507 &    0.960 &    71.8 &    70.3 &    73.1 &    10.24 &     0.25 &     1.04 &     0.22 &   -71.7 &     0.9 &   149 \myeol 
\hphantom{\knownbinary} & 23 &   398496965625177216 &      467 &      432 &      515 &    0.728 &    0.676 &    0.792 &   151.9 &   137.8 &   164.3 &    13.78 &     0.04 &     4.15 &     0.04 &  -151.4 &     7.9 &   106 \myeol 
\hphantom{\knownbinary} & 24 &  3007538204640292480 &    -4232 &    -4302 &    -4157 &    0.742 &    0.580 &    0.919 &    86.4 &    85.2 &    87.7 &     2.67 &     0.01 &     0.44 &     0.01 &    86.5 &     0.8 &   110 \myeol 
\hphantom{\knownbinary} & 25 &   911145876981562496 &     3660 &     3514 &     3818 &    0.749 &    0.503 &    1.014 &    31.2 &    29.9 &    32.3 &     8.61 &     0.09 &     0.82 &     0.09 &   -30.7 &     0.7 &    68 \myeol 
\hphantom{\knownbinary} & 26 &  2184252351930210560 &     5194 &     4172 &     7173 &    0.757 &    0.589 &    0.887 &    24.7 &    17.9 &    30.7 &     7.65 &     0.02 &     0.59 &     0.02 &   -24.1 &     3.9 &    66 \myeol 
\dubious\hphantom{\knownbinary} & 27 &  3320184202856435840 &       19 &       18 &       20 &    0.777 &    0.748 &    0.813 &   416.6 &   398.1 &   432.8 &   123.24 &     0.02 &  1026.46 &     0.01 &  -414.0 &    10.4 &    10 \myeol 
\hphantom{\knownbinary} & 28 &  6483100752169539584 &    -3837 &    -5038 &    -3006 &    0.787 &    0.435 &    1.066 &    34.9 &    26.8 &    44.3 &     7.35 &     0.04 &     1.49 &     0.04 &    34.4 &     5.4 &    36 \myeol 
\hphantom{\knownbinary} & 29 &  4155835025908320640 &      327 &      308 &      349 &    0.788 &    0.737 &    0.852 &   283.7 &   264.6 &   300.4 &    10.55 &     0.04 &     5.89 &     0.04 &  -283.0 &    10.7 &   107 \myeol 
\hphantom{\knownbinary} & 30 &  3443909634992792320 &    -3140 &    -3814 &    -2596 &    0.810 &    0.486 &    1.216 &    58.0 &    47.7 &    69.8 &     5.36 &     0.09 &     0.29 &     0.08 &    58.7 &     6.6 &   230 \myeol 
\hphantom{\knownbinary} & 31 &  6726602067616477056 &    -2057 &    -2074 &    -2038 &    0.815 &    0.790 &    0.841 &    26.8 &    26.6 &    27.1 &    17.70 &     0.03 &     1.79 &     0.03 &    26.9 &     0.1 &    56 \myeol 
\hphantom{\knownbinary} & 32 &  2926732831673735168 &    -1658 &    -1665 &    -1650 &    0.819 &    0.803 &    0.835 &    66.5 &    66.3 &    66.8 &     8.87 &     0.01 &     0.91 &     0.01 &    66.5 &     0.2 &   137 \myeol 
\hphantom{\knownbinary} & 33 &  1791617849154434688 &    -1533 &    -1545 &    -1520 &    0.837 &    0.819 &    0.854 &    55.9 &    55.5 &    56.3 &    11.42 &     0.02 &     1.23 &     0.01 &    55.8 &     0.3 &   109 \myeol 
\knownbinary & 34 &  3101404272522355584 &    -3728 &    -3781 &    -3672 &    0.843 &    0.651 &    1.043 &    97.2 &    96.5 &    98.0 &     2.70 &     0.02 &     0.23 &     0.02 &    97.4 &     0.4 &   242 \myeol 
\hphantom{\knownbinary} & 35 &  3218956015577464064 &    -3162 &    -3595 &    -2769 &    0.845 &    0.740 &    0.957 &    58.1 &    51.1 &    66.1 &     5.33 &     0.03 &     0.63 &     0.02 &    58.4 &     4.5 &   104 \myeol 
\hphantom{\knownbinary} & 36 &  2929487348824926336 &    -5238 &    -5338 &    -5128 &    0.845 &    0.519 &    1.167 &    70.8 &    69.9 &    71.8 &     2.64 &     0.02 &     0.45 &     0.02 &    70.7 &     0.6 &    88 \myeol 
\hphantom{\knownbinary} & 37 &  3621143693841328896 &    -1522 &    -1823 &    -1276 &    0.859 &    0.733 &    1.047 &    85.0 &    71.0 &   101.0 &     7.57 &     0.05 &     0.87 &     0.06 &    85.2 &     9.0 &   156 \myeol 
\hphantom{\knownbinary} & 38 &  3260079227925564160 &      929 &      912 &      947 &    0.862 &    0.846 &    0.878 &    32.8 &    32.1 &    33.3 &    32.13 &     0.03 &     6.14 &     0.03 &   -32.7 &     0.4 &    36 \myeol 
\dubious\hphantom{\knownbinary} & 39 &  3222258501830719616 &    -4375 &    -5066 &    -3774 &    0.863 &    0.299 &    1.531 &    80.0 &    69.3 &    92.2 &     2.80 &     0.05 &     0.73 &     0.04 &    80.4 &     6.9 &    65 \myeol 
\dubious\hphantom{\knownbinary} & 40 &   775766686745320960 &     -305 &     -351 &     -269 &    0.863 &    0.678 &    1.131 &   394.3 &   382.9 &   407.3 &     8.14 &     0.59 &     4.09 &     0.49 &   394.7 &     7.3 &   166 \myeol 
\hphantom{\knownbinary} & 41 &  5346007675222666752 &     -285 &     -287 &     -282 &    0.868 &    0.859 &    0.875 &    76.3 &    75.7 &    77.1 &    44.99 &     0.02 &    28.02 &     0.02 &    76.3 &     0.4 &    26 \myeol 
\dubious\knownbinary & 42 &  5473864079915092736 &     -137 &     -142 &     -132 &    0.876 &    0.842 &    0.909 &   711.6 &   689.6 &   736.7 &    10.04 &     0.05 &    12.92 &     0.05 &   712.5 &    14.1 &   117 \myeol 
\hphantom{\knownbinary} & 43 &   899893234465049216 &     1122 &      908 &     1529 &    0.879 &    0.709 &    1.194 &    41.5 &    30.4 &    51.3 &    20.99 &     0.04 &     3.80 &     0.03 &   -41.1 &     6.2 &    48 \myeol 
\hphantom{\knownbinary} & 44 &  5553958176239495040 &     -309 &     -330 &     -288 &    0.890 &    0.830 &    0.949 &    78.5 &    73.5 &    84.1 &    40.32 &     0.02 &    24.00 &     0.03 &    78.6 &     3.2 &    28 \myeol 
\dubious\hphantom{\knownbinary} & 45 &  4041250662947214464 &      755 &      667 &      868 &    0.896 &    0.422 &    1.452 &   162.7 &   150.8 &   173.2 &     7.99 &     0.57 &     2.09 &     0.72 &  -162.3 &     6.7 &   131 \myeol 
\hphantom{\knownbinary} & 46 &  2924378502398307840 &    -1871 &    -1889 &    -1851 &    0.899 &    0.880 &    0.919 &    85.6 &    84.8 &    86.4 &     6.11 &     0.01 &     0.76 &     0.01 &    85.5 &     0.5 &   145 \myeol 
\hphantom{\knownbinary} & 47 &  3972130276695660288 &     -522 &     -526 &     -517 &    0.900 &    0.891 &    0.908 &    31.2 &    30.9 &    31.5 &    59.96 &     0.03 &    21.84 &     0.03 &    31.1 &     0.2 &    18 \myeol 
\knownbinary & 48 &  1132123559268892288 &     2581 &     2473 &     2712 &    0.904 &    0.784 &    1.033 &    78.0 &    74.2 &    81.3 &     4.87 &     0.01 &     0.38 &     0.02 &   -77.3 &     2.1 &   210 \myeol 
\knownbinary & 49 &  6224087389269263488 &    -1722 &    -1793 &    -1647 &    0.909 &    0.830 &    0.996 &    18.0 &    17.3 &    18.8 &    31.47 &     0.15 &     3.28 &     0.17 &    18.0 &     0.4 &    36 \myeol 
\hphantom{\knownbinary} & 50 &  5551538941421122304 &    -3858 &    -4048 &    -3661 &    0.910 &    0.860 &    0.955 &    30.1 &    28.7 &    31.7 &     8.46 &     0.01 &     1.04 &     0.01 &    29.7 &     0.9 &    51 \myeol 
\knownbinary & 51 &  2933503521200215424 &    -4017 &    -4160 &    -3860 &    0.914 &    0.825 &    1.022 &    32.7 &    31.5 &    34.0 &     7.46 &     0.02 &     0.25 &     0.01 &    32.7 &     0.7 &   206 \myeol 
\hphantom{\knownbinary} & 52 &  2020810807491120896 &    -1560 &    -1936 &    -1270 &    0.922 &    0.257 &    2.050 &    76.8 &    68.8 &    85.8 &     8.19 &     0.82 &     0.84 &     0.66 &    77.0 &     5.1 &   159 \myeol 
\hphantom{\knownbinary} & 53 &  1926461164913660160 &       36 &       36 &       37 &    0.934 &    0.931 &    0.938 &    80.9 &    80.6 &    81.2 &   316.54 &     0.04 &  1595.62 &     0.03 &   -77.3 &     0.2 &     3 \myeol 
\hphantom{\knownbinary} & 54 &  3142271161216457472 &     -455 &     -496 &     -415 &    0.936 &    0.850 &    1.025 &   114.0 &   104.5 &   124.9 &    18.83 &     0.05 &     8.52 &     0.06 &   114.4 &     6.1 &    53 \myeol 
\hphantom{\knownbinary} & 55 &  5896469620419457536 &     5479 &     4940 &     6241 &    0.956 &    0.815 &    1.047 &    20.6 &    18.0 &    22.8 &     8.68 &     0.02 &     0.78 &     0.01 &   -20.5 &     1.4 &    48 \myeol 
\hphantom{\knownbinary} & 56 &  5853498713190525696 &       27 &       27 &       27 &    0.958 &    0.951 &    0.965 &    32.4 &    32.1 &    32.6 &   768.09 &     0.05 &  3859.23 &     0.03 &   -21.9 &     0.2 &     1 \myeol 
\hphantom{\knownbinary} & 57 &  1726458694148645888 &     5583 &     4869 &     6666 &    0.965 &    0.333 &    2.667 &    37.5 &    31.5 &    42.8 &     4.73 &     0.02 &     1.14 &     0.02 &   -35.9 &     3.5 &    31 \myeol 
\dubious\hphantom{\knownbinary} & 58 &   418338821185634048 &     3098 &     2654 &     3813 &    0.974 &    0.834 &    1.172 &    24.2 &    19.6 &    28.3 &    13.04 &     0.04 &     0.94 &     0.04 &   -24.0 &     2.6 &    70 \myeol 
\knownbinary & 59 &  6899603831309106560 &     2460 &     2156 &     2921 &    0.977 &    0.840 &    1.181 &    53.0 &    44.6 &    60.4 &     7.51 &     0.03 &     0.92 &     0.02 &   -52.4 &     4.7 &    90 \myeol 
\hphantom{\knownbinary} & 60 &  3676827188919831936 &     7335 &     7202 &     7490 &    0.991 &    0.870 &    1.105 &    13.9 &    13.6 &    14.1 &    10.10 &     0.02 &     1.63 &     0.02 &   -11.8 &     0.2 &    15 \myeol 
\dubious\hphantom{\knownbinary} & 61 &  5614610776700908672 &     -718 &     -793 &     -654 &    0.998 &    0.823 &    1.206 &    88.4 &    81.9 &    95.8 &    15.38 &     0.49 &     4.50 &     0.30 &    88.7 &     4.2 &    64 \myeol 

\bottomrule
\end{tabular}
}
\end{table*}

Table~\ref{tab:periStats} lists the encounter parameters as well as the parallaxes, proper motions, and radial velocities of the 61 closest encounters. 
The final column of the table gives the ratio of the current absolute radial velocity (\rv) to the current transverse velocity (\vt) of each star. The fact that these ratios are generally much larger than 1 is a selection effect: stars that approach close to the Sun are those that are currently moving almost directly toward or away from it, so have $|\rv| >> \vt$.
It is therefore not surprising that many of the close encounters have large absolute radial velocities. Nonetheless, we see several encounters with radial velocities (and thus space velocities) above several hundred \kms. While such velocities do occur, we should be suspicious of these, as they could be a result of an incorrect template or failure of the cross-correlation in the \gaia\ processing reported in \cite{2022arXiv220605902K}. This will be discussed for some individual cases in section~\ref{sec:results:specific}. 

\begin{table*}[th]
\centering
\tiny{
\caption{
Additional data on the close encounters listed in Table~\ref{tab:periStats}.
All fields are taken directly from the \gdr{3} {\tt gaia\_source} table, except for \mg\ which is  $\gmag + 5\log_{10}(\parallax/100)$, where \parallax\ is the zeropoint-corrected parallax in mas.
Descriptions of the other fields can be found in the \href{https://gea.esac.esa.int/archive/documentation/GDR3/Gaia_archive/chap_datamodel/sec_dm_main_source_catalogue/ssec_dm_gaia_source.html}{\gdr{3} online documentation}.
\label{tab:extras}
}
\tabcolsep=0.14cm
\begin{tabular}{*{13}{r}}
\toprule
& \gdr{3} {\tt source\_id} & \gmag & \bprp & \mg    & {\tt ruwe} & {\tt astrometric\_}  & {\tt ipd\_frac\_}  &  {\tt rv\_ }                 & {\tt rv\_nb\_}   & $l$ & $b$ \\
&                                   & mag      & mag    &   mag &           & {\tt params\_}         &  {\tt multi\_}          &  {\tt expected\_}    & {\tt transits}  & deg & deg \\
&                                   &                &             &             &           & {\tt solved}             & {\tt peak}               & {\tt sig\_to\_noise}  &                         &         &         \\
\midrule
 1 &  4270814637616488064 &  9.06 &  1.69 &  7.66 &   0.91 & 31 &   0 & 132.6 &  6 &  28 &   6 \myeol 
 2 &   510911618569239040 &  8.89 &  0.76 &  4.50 &   2.01 & 31 &   0 & 188.8 & 18 & 126 &  -1 \myeol 
 3 &  5544743925212648320 & 14.35 &  0.68 & 14.11 &   1.04 & 31 &   0 &   6.2 & 16 & 253 &  -1 \myeol 
 4 &   213090546082530816 & 12.06 &  0.99 &  5.58 &   1.05 & 31 &   0 &  38.8 & 25 & 161 &   5 \myeol 
 5 &  5571232118090082816 & 11.78 &  1.49 &  6.83 &   0.93 & 31 &  26 &  31.2 &  9 & 250 & -25 \myeol 
 6 &  5469802896279029504 & 10.08 &  1.62 &  7.42 &   1.30 & 31 &   0 & 119.3 & 13 & 271 &  29 \myeol 
 7 &  3372104035275483392 & 15.31 &  2.57 & 10.70 &   0.97 & 31 &   0 &   3.7 &  8 & 194 &   4 \myeol 
 8 &  3207963476278403200 & 14.89 &  2.63 &  7.69 &   1.08 & 95 &  28 &   4.2 &  9 & 208 & -25 \myeol 
 9 &  3106500096597409792 & 12.79 &  0.89 &  5.22 &   0.99 & 31 &   0 &  15.9 & 11 & 214 &  -3 \myeol 
10 &  4763293626627587840 & 16.00 &  2.75 & 10.19 &   1.24 & 95 &  14 &   3.4 & 17 & 266 & -34 \myeol 
11 &   929788371508812288 & 15.83 &  3.20 & 11.52 &   1.10 & 31 &   2 &   2.3 &  6 & 173 &  32 \myeol 
12 &  6913732624445112832 & 14.53 &  1.38 &  7.04 &   1.01 & 31 &   0 &   7.8 & 33 &  43 & -30 \myeol 
13 &  3118526069444386944 & 12.14 &  1.61 &  6.55 &   3.19 & 31 &   0 &  30.5 & 12 & 210 &  -7 \myeol 
14 &  6608946489396474752 & 12.27 &  1.42 &  6.76 &   1.07 & 31 &   0 &  31.5 & 15 &  24 & -61 \myeol 
15 &  3295253979286613376 & 14.33 &  2.35 &  9.18 &   1.09 & 31 &   0 &  15.3 & 28 & 188 & -19 \myeol 
16 &  1952802469918554368 & 10.83 &  2.81 & 11.59 &   1.19 & 31 &   8 & 145.0 & 25 &  88 & -12 \myeol 
17 &  5261593808165974784 & 12.68 &  2.02 &  8.61 &   1.14 & 31 &   0 &  38.7 & 25 & 286 & -27 \myeol 
18 &  3054509410098672000 & 12.40 &  2.56 &  9.39 &   1.43 & 31 &   1 &  51.5 & 21 & 224 &   6 \myeol 
19 &  3001468183198140800 & 15.24 &  2.08 &  8.51 &   1.07 & 31 &   0 &   4.2 &  7 & 219 & -12 \myeol 
20 &  4116451378388951424 & 15.80 &  1.90 & 11.17 &   3.58 & 95 &  74 &   2.2 &  6 &   4 &   4 \myeol 
21 &  3600338081985998080 & 14.19 &  2.81 & 11.26 &   3.24 & 31 &   0 &  18.1 & 23 & 275 &  56 \myeol 
22 &  4536673181955253504 & 13.47 &  2.32 &  8.51 &  20.37 & 31 &   0 &  27.6 & 32 &  53 &  16 \myeol 
23 &   398496965625177216 & 15.45 &  2.85 & 11.14 &   1.06 & 31 &   0 &   3.3 &  2 & 128 & -15 \myeol 
24 &  3007538204640292480 & 12.36 &  0.82 &  4.48 &   1.09 & 31 &   0 &  24.1 & 16 & 216 & -10 \myeol 
25 &   911145876981562496 & 12.48 &  1.50 &  7.15 &   6.28 & 31 &   0 &  32.1 & 16 & 184 &  35 \myeol 
26 &  2184252351930210560 & 15.16 &  2.47 &  9.56 &   1.07 & 31 &   0 &   6.9 & 21 &  87 &   9 \myeol 
27 &  3320184202856435840 & 13.97 &  0.83 & 14.42 &   0.92 & 95 &   0 &   6.5 & 12 & 202 & -10 \myeol 
28 &  6483100752169539584 & 15.52 &  2.45 &  9.84 &   1.04 & 31 &   0 &   3.9 & 19 & 353 & -42 \myeol 
29 &  4155835025908320640 & 15.26 &  2.75 & 10.37 &   0.97 & 31 &   0 &   2.9 &  4 &  22 &   1 \myeol 
30 &  3443909634992792320 & 14.04 &  1.69 &  7.67 &   2.58 & 95 &  99 &   6.8 &  4 & 180 &   2 \myeol 
31 &  6726602067616477056 &  9.18 &  0.98 &  5.42 &   1.23 & 31 &  17 &  54.7 &  2 & 354 & -12 \myeol 
32 &  2926732831673735168 &  9.56 &  0.71 &  4.30 &   1.02 & 31 &   0 & 148.2 & 22 & 231 & -12 \myeol 
33 &  1791617849154434688 & 11.00 &  1.08 &  6.28 &   1.01 & 31 &   0 &  80.5 & 23 &  71 & -18 \myeol 
34 &  3101404272522355584 & 11.22 &  0.67 &  3.34 &   1.12 & 31 &   0 &  31.3 &  9 & 219 &  -0 \myeol 
35 &  3218956015577464064 & 14.85 &  2.02 &  8.47 &   1.00 & 31 &   0 &   6.7 & 24 & 206 & -16 \myeol 
36 &  2929487348824926336 & 11.22 &  0.77 &  3.30 &   1.30 & 95 &   0 &  47.5 & 17 & 233 &  -5 \myeol 
37 &  3621143693841328896 & 15.55 &  2.29 &  9.93 &   1.16 & 31 &   0 &   3.6 & 11 & 310 &  49 \myeol 
38 &  3260079227925564160 & 11.73 &  2.13 &  9.27 &   1.62 & 31 &   0 &  64.1 & 16 & 189 & -34 \myeol 
39 &  3222258501830719616 & 15.16 &  2.51 &  7.35 &   1.79 & 31 &   0 &   8.4 & 20 & 201 & -18 \myeol 
40 &   775766686745320960 & 15.91 &  2.91 & 10.45 &  13.85 & 31 &  15 &   3.6 &  8 & 183 &  62 \myeol 
41 &  5346007675222666752 & 12.46 &  2.69 & 10.72 &   1.19 & 31 &   0 &  53.3 & 23 & 290 &   5 \myeol 
42 &  5473864079915092736 & 15.64 &  2.93 & 10.64 &   1.42 & 31 &   0 &   4.0 & 12 & 263 &  27 \myeol 
43 &   899893234465049216 & 14.96 &  2.72 & 11.56 &   1.15 & 31 &   0 &   4.3 &  4 & 180 &  22 \myeol 
44 &  5553958176239495040 & 14.79 &  3.29 & 12.82 &   1.23 & 31 &   0 &  10.6 & 24 & 255 & -24 \myeol 
45 &  4041250662947214464 & 16.26 &  2.54 & 10.76 &  12.32 & 95 &  52 &   5.1 & 13 & 354 &  -3 \myeol 
46 &  2924378502398307840 & 12.61 &  1.23 &  6.53 &   0.99 & 31 &   0 &  39.2 & 38 & 232 & -16 \myeol 
47 &  3972130276695660288 &  9.88 &  2.17 &  8.77 &   1.18 & 31 &   0 & 143.4 & 12 & 228 &  66 \myeol 
48 &  1132123559268892288 & 13.64 &  1.41 &  7.06 &   0.95 & 31 &   0 &  11.4 & 18 & 133 &  34 \myeol 
49 &  6224087389269263488 & 10.64 &  1.85 &  8.13 &   8.44 & 31 &  32 &  59.7 &  6 & 334 &  26 \myeol 
50 &  5551538941421122304 & 13.08 &  1.70 &  7.71 &   1.02 & 31 &   0 &  26.4 & 19 & 258 & -22 \myeol 
51 &  2933503521200215424 &  8.79 &  0.51 &  3.14 &   0.89 & 31 &   1 & 189.1 & 19 & 230 &  -9 \myeol 
52 &  2020810807491120896 & 14.79 &  1.96 &  9.34 &   8.07 & 95 &  30 &   9.2 & 33 &  61 &   1 \myeol 
53 &  1926461164913660160 & 10.38 &  3.53 & 12.88 &   1.03 & 31 &   8 & 142.5 & 12 & 110 & -17 \myeol 
54 &  3142271161216457472 & 15.75 &  3.13 & 12.12 &   1.14 & 31 &   0 &   3.7 & 11 & 211 &  13 \myeol 
55 &  5896469620419457536 & 13.53 &  1.98 &  8.21 &   0.97 & 31 &   1 &  17.0 & 24 & 314 &   7 \myeol 
56 &  5853498713190525696 &  8.98 &  3.80 & 13.41 &   0.97 & 95 &  11 & 222.3 &  7 & 314 &  -2 \myeol 
57 &  1726458694148645888 & 14.65 &  1.95 &  8.01 &   1.03 & 31 &   0 &   8.2 & 23 & 124 &  32 \myeol 
58 &   418338821185634048 & 14.76 &  2.87 & 10.33 &   2.12 & 31 &   0 &  11.1 & 30 & 122 &  -8 \myeol 
59 &  6899603831309106560 & 14.76 &  2.15 &  9.13 &   1.16 & 31 &   0 &   7.6 & 18 &  32 & -29 \myeol 
60 &  3676827188919831936 & 10.12 &  0.88 &  5.14 &   0.97 & 31 &   0 & 134.2 & 31 & 302 &  57 \myeol 
61 &  5614610776700908672 & 14.36 &  1.05 & 10.29 &  29.74 & 95 &  27 &   8.1 & 41 & 241 &  -0 \myeol 

\bottomrule
\end{tabular}
}
\end{table*}

Table~\ref{tab:extras} provides additional information on the encounters.
\mg\ is the absolute magnitude of the source on the assumption of zero extinction and $1/\parallax$ being a reasonable distance estimate, which it is because the parallax SNR is above 30 for all but three of the sources (and those still have SNRs above 10). 
{\tt ruwe} is a calibrated reduced $\chi^2$ of the astrometric solution: values much more than a few could indicate the astrometric solution is poor, although this is not a definitive metric.
{\tt astrometric\_params\_solved} is 31 if the astrometric solution was for the standard five parameters (2 positions, parallax, 2 proper motions), and 95 if the colour of the source was solved for in addition, for reasons discussed in \cite{2021A&A...649A...2L}.
{\tt ipd\_frac\_multi\_peak} indicates the percentage of image windows used in the astrometry which contain more than one peak. This may indicate it is a double star (physical or otherwise), although only two of the stars in the table have a definite indication of physical close binarity in \gdr{3} (rows 13 and 58).
{\tt rv\_expected\_sig\_to\_noise} is the estimated SNR of the spectrum used to determine the RV, and {\tt rv\_nb\_transits} is the number of measurement epochs used in that determination.
$l$ and $b$ are the Galactic longitude and latitude of the star.

\begin{figure*}[th]
\begin{center}
\includegraphics[width=0.75\textwidth, angle=0]{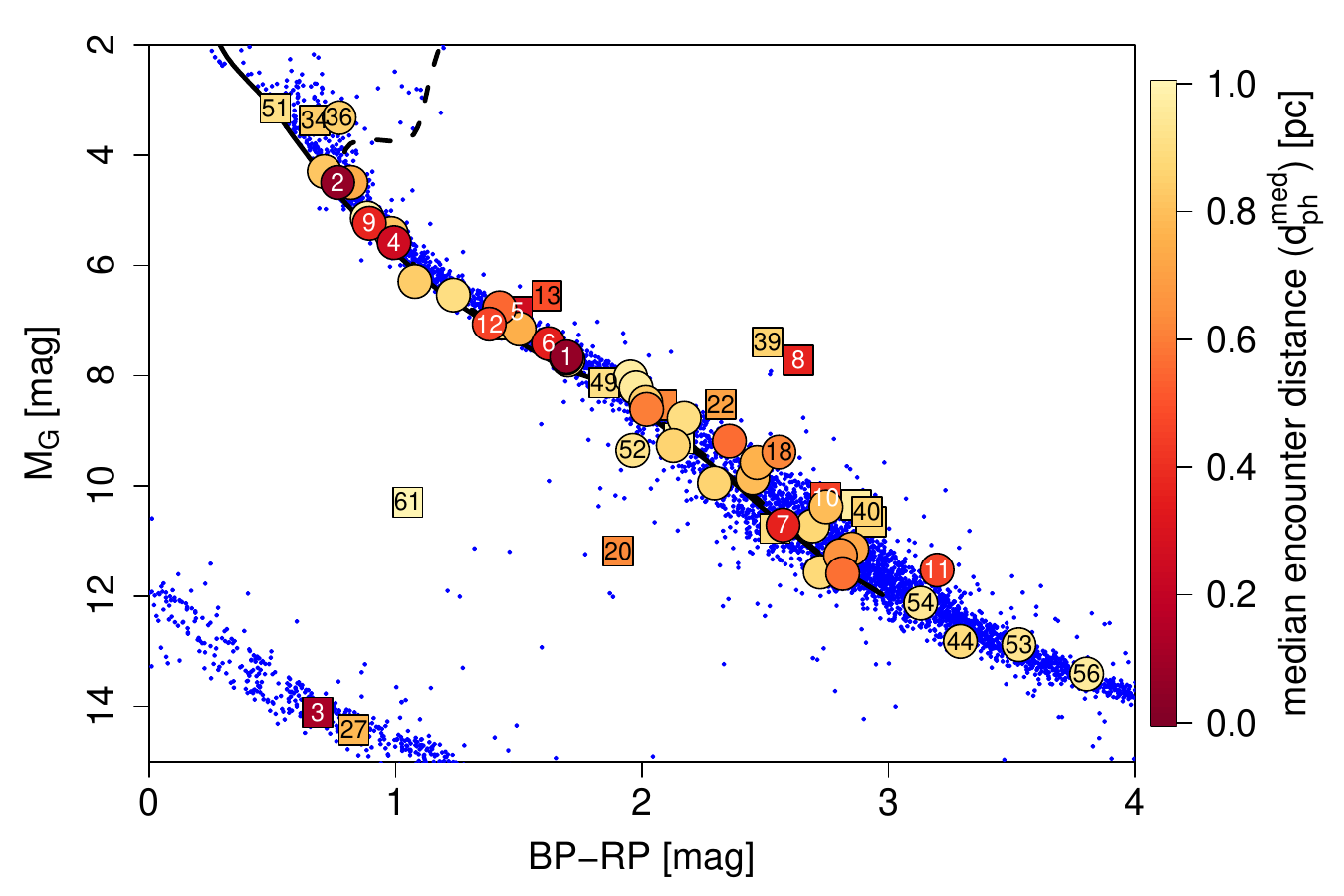}
\caption{Colour--absolute magnitude diagram, with \mg\ computed assuming zero extinction. The red/yellow coloured points are the 61 encounters with $\dphmed<1$\,pc identified in this work, colour-coded according to \dphmed. 
Circles denote good encounters; squares denote the questionable encounters, which are marked with a \dubious\ or \knownbinary\ in Table~\ref{tab:periStats}.
The 13 encounters with $\dphmed<0.5$\,pc plus several others of possible interest are labelled by their order of increasing distance (row number in Table~\ref{tab:periStats}).
For orientation, the black lines are unreddened solar metallicity PARSEC isochrones for 1\,Gyr (solid) and 10\,Gyr (dashed) from \cite{2017ApJ...835...77M}, 
and the small blue dots are a random subset of all sources in \gdr{3} with $\parallax>50$\,mas and {\tt ruwe}\,$<1.2$.
\label{fig:qcd}
}
\end{center}
\end{figure*} 

Figure~\ref{fig:qcd} shows the location of encounters in the colour--absolute magnitude diagram. The median distance to these stars is 100\,pc (maximum 380\,pc), so in most cases the assumption of negligible extinction is valid. The location of the stars in this diagram, as well as the 
stellar parameters provided in \gdr{3} for many of them by \cite{2022arXiv220606138A}, show that most of the stars are main sequence FGKM stars. In addition there are two white dwarfs, two possible very young stars lying above the main sequence, and two stars lying unexpectedly between the white dwarf and main sequences. These will be discussed in the next section.

Encounters that are dubious are prefixed with a \dubious\ in Table~\ref{tab:periStats}. This assignment occurs because the RV is suspiciously large, the {\tt ruwe} is large, or for another reason given in the next section. Eight stars that are probably in wide binaries according to the criteria of \cite{2021MNRAS.506.2269E} are prefixed with a ``b". Unless the orbital period is very long compared to the time to encounter (\tphmed), the orbital motion will be significant, and hence the Galactic trajectory computed in the present work will be incorrect.  
\cite{2021MNRAS.506.2269E} does not list periods, and only projected separations (which range from $10^2$ to $10^4$\,au for these eight stars), so until a long period can be confirmed for any of these stars, their solar encounters should be regarded as unreliable.
Two other stars in Table~\ref{tab:periStats} (6608946489396474752 and  3001468183198140800) are also listed in this wide binary catalogue, but they are probably chance alignments (values of {\tt R\_chance\_align} of 0.44 and 2.3 respectively, whereas the ones denoted ``b" have 0.07 or -- generally -- much less).

\subsection{Specific close encounters}\label{sec:results:specific}

Here I discuss selected sources from Table~\ref{tab:periStats} giving the row number, \gaia\ {\tt source\_id}, and the primary Simbad name for the source, if available. The {\tt source\_id} has not changed from EDR3 to DR3 (other than this prefix).

\begin{description}

\item[1]  \object{Gaia DR3 4270814637616488064} = \object{Gl 710}\\
Since Hipparcos, this K7 dwarf has often held the title of the closest 
known bona-fide encounter.
In \citetalias{2015A&A...575A..35B} (Hipparcos) it had a median encounter distance of
0.267\,pc with a \CI\ of 0.101--0.444\,pc.
In \citetalias{2018A&A...616A..37B} (\gdr{2}) this was significantly reduced to 0.0676\,pc (\CI\ 0.0519--0.0842\,pc) where it was the closest encounter found.
This does not change much using \gdr{3} -- the median is 0.0636\,pc --  but it is more precise with a \CI\ of 0.0595--0.0678\,pc. This is mostly due to a factor of three improvement in the precision of both its parallax and proper motions. 
Assuming Gl 710 has a mass of 0.7\,\Msol\
(\citetalias{2018A&A...616A..37B})
the gravitational attraction by the Sun only lowers the median perihelion by 7\,au 
(35\,$\mu$pc, 0.06\%)
demonstrating that gravitational focusing can be neglected in all these encounters.

\item[2] \object{Gaia DR3 510911618569239040} = \object{HD 7977}\\
This G3 dwarf was found in \gdr{2} \citepalias{2018A&A...616A..37B} as a close encounter with a median distance of 
0.429\,pc (\CI\ 0.368--0.494\,pc), but now comes much closer
with a median of 0.0641\,pc (\CI\ 0.0191--0.1171\,pc).
This is mostly due to a significant reduction in its proper motion.
There is a large relative uncertainty in the encounter distance because the SNR of the proper motion is low, just 4.2.  Yet this is the only known star with a significant probability -- 32\% -- of approaching within 0.05\,pc of the Sun.
The {\tt ruwe} is slightly inflated (2.01), which could suggest a problem with the astrometric solution, although there is no indication of close binarity.
\cite{2022arXiv220611047D} found a much closer minimum perihelion distance for this star using \gaia~EDR3 (which has the same astrometry as DR3, but different RVs), but this is partly because they averaged the {\em signed} coordinates of the surrogates (section~\ref{sec:method}), which underestimates the distance in this case (see their Figure 5).

\item[3] \object{Gaia DR3 5544743925212648320} = \object{UPM J0812-3529}\\
This was discovered to be a nearby white dwarf by \cite{2018AJ....155..176F}, but \gdr{3} is the first publication of a radial velocity.  If a real encounter, then it will pass just 0.11\,pc from the Sun in 29\,kyr.
However, the radial velocity is suspiciously large ($-374$\,\kms), even though the ratio to the transverse velocity is modest (it has a large proper motion and is currently just 11.2\,pc away). It is hard to measure radial velocities for white dwarfs accurately on account of their relatively featureless spectra, especially in this case due to its apparently strong magnetic fields \citep{2020A&A...643A.134B}. The \gdr{3} RV pipeline does not include any white dwarf templates, and although the main sequence dwarf template \teff\ of 6000\,K is close to the value of 6090\,K estimated by 
\cite{2020A&A...643A.134B}, the \gdr{3} RV is likely wrong (\gaia\ CU6 (RV) team, private communication). If it is correct, the strong gravitational redshift in white dwarfs means that the true RV of the star would be tens of \kms\ more negative. As the RV of this star is negative, the true absolute value of its RV would be even larger and the encounter even closer. If the gravitational redshift were 30\,\kms, for example, the star would come about 0.008\,pc (7\%) closer.

\item[4] \object{Gaia DR3 213090546082530816}=\object{UCAC4 689-035468}\\
This was found in \citetalias{2018A&A...616A..37B} but with a much larger median encounter distance of
1.44\,pc and large uncertainty (\CI) of 0.703--2.175\,pc. These have now decreased to 
0.248\,pc and 0.072--0.489\,pc respectively thanks to more precise astrometry. 

\item[6] \object{Gaia DR3 5469802896279029504} = \object{CD-25 8217}\\
At $11\pm 1$\,Myr, this is the latest encounter. It is also the slowest encounter, moving at just 3.2\,\kms\ relative to the Sun. Both of these are a consequence of its small radial velocity of $-2.66\pm0.17$\,\kms.
RAVE DR6 \citep{2020AJ....160...82S} reports an RV of
$-4.02\pm1.06$\,\kms. If we use this in the orbital integration we get a larger median perihelion distance of 0.92\,pc with a wider \CI\ of 0.32--1.10\,pc, and an encounter that probably takes places sooner, at 7.7\,Myr, but with a larger spread due to the larger RV uncertainty (\CI\ 5.6--12.9\,Myr).

\item[8] \object{Gaia DR3 3207963476278403200}\\
This has a suspiciously high RV of $515\pm6$\,\kms.
There is a second source in \gdr{3} 1.0$"$ away that is 1.1\,mag fainter;
this could have interfered with the radial velocity determination. The RV spectrum also has a very slow SNR.
The star also lies well above the main sequence (Figure~\ref{fig:qcd}).
This can indicate binarity, although 3 magnitudes offset is far too large.
It could instead be a pre-main sequence star (see object 39 below), in which case
the RV is likely wrong due to a lack of appropriate templates in the data processing.

\item[13] \object{Gaia DR3 3118526069444386944}\\
This was identified as a young stellar object candidate by \cite{2018A&A...620A.172Z}.
It was found in \citetalias{2018A&A...616A..37B}  to encounter at 1.03\,pc; now apparently reduced to 0.49\,pc. However, it
is just one of two stars in Table~\ref{tab:periStats} that is identified as a close physical binary in \gdr{3} via the {\tt non\_single\_star} flag (the other is object 58).
Specifically, it is an astrometric binary with a significant proper motion acceleration (a so-called 7-parameter solution). The table {\tt nss\_acceleration\_astro} lists improved parallaxes and proper motions that take into account this acceleration, but as they still may not be representative of the centre-of-mass of the binary system, I have not recomputed the encounter. This encounter should be disregarded until the system motion has been adequately characterized.

\item[20] \object{Gaia DR3 4116451378388951424}\\
This star is strange because it lies in the otherwise relatively empty region between the white dwarf and main sequences. After \gdr{2} it was found that many stars in this region were very faint ($\gmag \gtrsim 19.5$\,mag) and had spuriously large parallaxes. This situation has been greatly improved in \gdr{3}, but as this star has a large {\tt ruwe}, its astrometric solution may well be inappropriate.
It lies very close to the Galactic centre in a crowded field, so it is possible that a nearby star has affected the measurement. This should lead us to doubt the surprisingly large RV too (which is extracted from a very low SNR spectrum).

\item[27] \object{Gaia DR3 3320184202856435840} = \object{EGGR 290}\\
The apparent close encounter of this white dwarf in just 19\,kyr is likely erroneous on account of its unreliable large RV (see object 3 above).

\item[39] \object{Gaia DR3 3222258501830719616} = \object{ CVSO 29}\\
This is a T Tauri star in the Orion OB1 association. 
The fact that it is still contracting towards the main sequence explains its position above the main sequence in Figure~\ref{fig:qcd}, although any circumstellar dust extinction and redenning has not been taken into account in that plot.
This encounter supposedly occurred 4.4\,Myr ago, but the star probably had not even formed at that time. 
It is quite likely that interactions in its birth cluster have modified its motion. This encounter must be disregarded.

\item[53] \object{Gaia DR3 1926461164913660160} =
 \object{Ross 248}\\
This is currently one of the closest stars to the Sun at 3.2\,pc. It will come closer in 36\,kyr, to 0.934\,pc. 7\,kyr before this it will pass within 0.46\,pc of the interstellar object 'Oumuamua \citep{2018AJ....156..205B}.

\item[56] \object{Gaia DR3 5853498713190525696} = \object{Proxima Cen.}\\
It appears that our closest neighbour will get slightly closer in 27\,kyr, although its orbit around Alpha Centauri AB needs to be taken into account.
This was done by \cite{2020A&A...640A.129W}. They infer similar encounter parameters to the medians reported in Table~\ref{tab:periStats}, which is not surprising because Proxima Centauri's orbital period of $547^{+66}_{-40}$\,kyr
\citep{2017A&A...598L...7K}
is much longer than the time to perihelion of 
26.622\,kyr (\CI\ 26.580--26.653\,kyr).

\end{description}

\section{Conclusions} \label{sec:summary}

This study has revealed 61 stars that, based on \gdr{3} data, have passed -- or will pass -- within 1\,pc of the Sun, most within $\pm 6$\,Myr. Closer inspection reveals that 12 of these are probably not real encounters due to spurious measurements or being in close binaries. A further seven are in wide binaries, so their encounters are probably spurious unless their orbital periods are very long.
As \gdr{3} has much better precision and accuracy than previous \gaia\ releases, encounters within 1\,pc identified using earlier \gaia\ releases that do not appear in Table~\ref{tab:periStats} should generally be considered as obsolete.  Potential exceptions are multiple systems in which the orbit of the centre-of-mass has been  traced, and stars with reliable non-Gaia astrometry.

In principle one could use these results to infer an overall stellar encounter rate, as was done in \cite{2018A&A...609A...8B} and \cite{2018A&A...616A..37B}.
Yet correcting for the significant and poorly characterized incompleteness of the sample makes this difficult, and the result uncertain. This will be tackled using the larger sample expected in \gdr{4}. 
As close encounters generally have large ratios of radial velocities to transverse motions, contamination from spuriously large RVs is a problem. Resolved binaries not identified as such are also a source of contamination.
But the individual encounters identified in this paper can be used to reassess Oort cloud comet perturbation or effects on specific comets (e.g.\ \citealt{2022A&A...660A.100D}).

To extend the identification of close encounters significantly beyond $\pm 6$\,Myr we need to obtain RVs for fainter stars. 
There will be perhaps 100 million RVs in \gdr{4} \citep{2022arXiv220605902K}, and many will be near the faint limit of the RV survey, which is expected to extend two magnitudes deeper from the current limit of $G_{\rm RVS}=14.0$\,mag to $G_{\rm RVS}=16.0$\,mag (the bright limit in \gdr{3} is $G_{\rm RVS}=2.8$\,mag).
Although one can estimate an RV (or rather a probability distribution over the RV) for  stars using their distances, transverse velocities, and a Galaxy model, and this will help to infer the statistics of encounters for fainter stars, this cannot be used to identify individual encounters reliably.

Whilst we need more RVs, more precise proper motions will help to reduce the uncertainties on the encounter distances. This is because the closest encountering stars tend to have very low proper motions, and the lower the proper motion, the lower its uncertainty must be to retain a given SNR. Yet as proper motion precision improves as the $3/2$ power of the observation baseline duration, and \gdr{3} is based on just under 3 years of observations, we can expect significant improvements in \gdr{4} (5 years) and \gdr{5} (expected to be 10.5--11 years). The detection of astrometric orbits with \gaia\ will also help include or remove multiple stellar systems from the encounter lists.

\begin{acknowledgments}
I would like to thank Yves Fr\'emat, Paola Sartoretti, and David Katz of the \gaia\ CU6 team, and Kareem El-Badry, for comments on the (un)reliability of white dwarf radial velocities. Thanks go also to the anonymous referee for useful suggestions.
This work is based on data from the European Space Agency (ESA) mission \gaia\ (\url{https://www.cosmos.esa.int/gaia}), processed by the \gaia\ Data Processing and Analysis Consortium (DPAC, \url{https://www.cosmos.esa.int/web/gaia/dpac/consortium}). Funding for the DPAC has been provided by national institutions, in particular the institutions participating in the \gaia\ Multilateral Agreement.
This research has made use of
the Simbad object database \citep{2000AAS..143....9W},
the VizieR catalogue access tool \citep{2000A&AS..143...23O},
and the Aladin sky atlas \citep{2014ASPC..485..277B} (all operated at \href{http://cds.u-strasbg.fr/}{CDS} Strasbourg),
as well as NASA’s Astrophysics Data System.
\end{acknowledgments}

\bibliography{bibliography}{}
\bibliographystyle{aasjournal}

\end{document}